\documentclass{PoS}

\usepackage{physics}
\usepackage{enumitem}

\title{Computing general observables in lattice models with complex actions}

\ShortTitle{Computing general observables in lattice models with complex actions}

\author{
	\speaker{Olmo Francesconi},$^{ab}$\thanks{This work has been partially supported by the ANR project ANR-15-IDEX-02}
	Markus Holzmann,$^{a}$\thanks{Partially supported by Fondation NanoSciences (Grenoble)}
	Biagio Lucini,$^{c}$\thanks{Supported in part by the Royal Society Wolfson Research Merit Award WM170010 and by the STFC Consolidated Grant ST/P00055X/1 and by the European Research Council (ERC) under the European Union's Horizon 2020 research and innovation programme under grant agreement No 813942}
	Antonio Rago,$^{d}$\thanks{Supported by the STFC Consolidated Grant ST/P000479/1}
	and Jarno Rantaharju$^{e}$\thanks{Supported by the Academy of Finland grants 320123 and 308791}\\
	\llap{$^a$}Univ. Grenoble Alpes, CNRS, LPMMC, 38000 Grenoble, France\\
	\llap{$^b$}Physics Department, College of Science, Swansea University (Singleton Campus), Swansea SA2 8PP, UK\\ 
	\llap{$^c$}Department of Mathematics, Computational Foundry, Bay Campus, Swansea University, Swansea SA1 8EN, UK\\
	\llap{$^d$}Centre for Mathematical Sciences, University of Plymouth, Plymouth, PL4 8AA, UK\\
	\llap{$^e$}Department of Physics \& Helsinki Institute of Physics, P.O. Box 64,FI-00014 University of Helsinki, Finland\\
	E-mail:	\email{o.francesconi.961603@swansea.ac.uk}, 
			\email{markus.holzmann@grenoble.cnrs.fr}, 
			\email{b.lucini@swansea.ac.uk}, 
			\email{antonio.rago@plymouth.ac.uk}, 
			\email{jarno.rantaharju@helsinki.fi}}

\abstract{The study of QFTs at finite density is hindered by the presence of the so-called sign problem. The action definition of such systems is, in fact, complex-valued making standard importance sampling Monte Carlo methods ineffective. In this work, we shall review the generalized density of states method for complex action systems and the Linear Logarithmic Relaxation algorithm (LLR). We will focus on the recent developments regarding the bias control of the LLR method and the evaluation of general observables in the DoS+LLR framework. Recent results on the well-known relativistic Bose gas will be presented, proving that in our approach the phase factor can be consistently evaluated over hundreds of orders of magnitude. A first exploratory study on the Thirring model in the DoS formalism will be presented as well.
}

\FullConference{37th International Symposium on Lattice Field Theory - Lattice2019\\
		16-22 June 2019\\
		Wuhan, China}

\begin{document}

\section{Introduction}
\begin{figure}[b]
\centering
\includegraphics[width=0.49\textwidth]{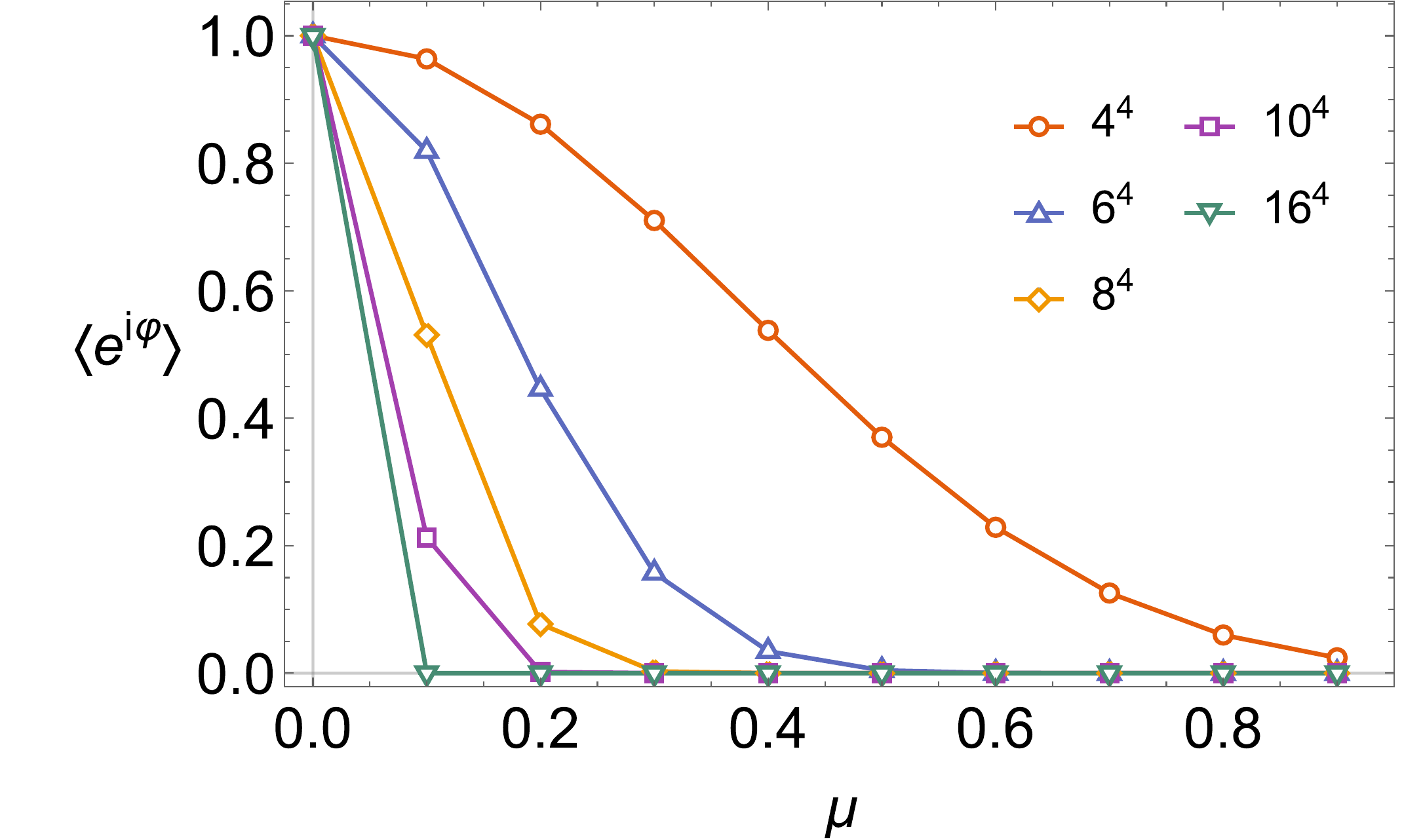}
\includegraphics[width=0.49\textwidth]{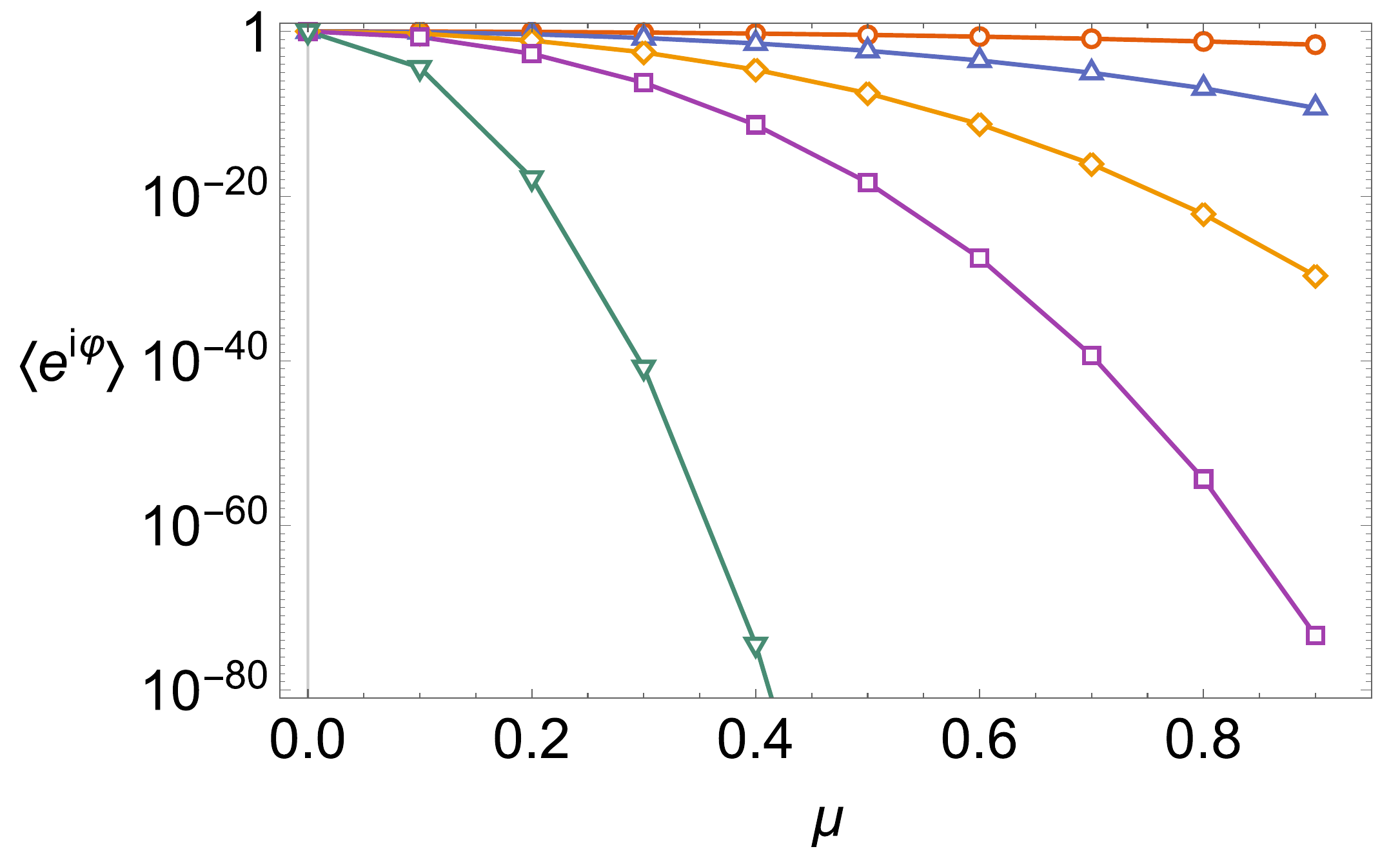}
\caption{Phase factor of the relativistic Bose gas for values of the chemical potential ranging from 0 to 0.9 and volumes $4^4$, $6^4$, $8^4$,$10^4$ and $16^4$. \textbf{Left}: result from a phase quenched simulation. \textbf{Right}: results from the DoS approach using the LLR method.}
\label{fig:ph_comp}
\end{figure}
A wide range of systems at finite density are described by partition functions that can be written in the form
\begin{equation} \label{eq:part}
Z(\mu)=\int {\cal D}\phi \ e^{-S^R[\phi]-i\mu S^I[\phi]}.
\end{equation}
In this expression, we have separated the real and imaginary part of the action and assumed the latter to be coupled linearly by $\mu$ for simplicity, but any function of $\mu$ can be assumed. At $\mu=0$ the integrand of Eq.\eqref{eq:part} can be interpreted as a Boltzmann weight and standard Monte Carlo techniques can be used in numerical studies. At finite values of the chemical potential the weight becomes complex leading to the failure of direct Monte Carlo methods. This is known as the \textit{sign problem} (see \cite{Gattringer:2016kco} for a recent review).

In our contribution we will present the results recently discussed in \cite{Francesconi:2019nph}, further developing the density of states method (originally proposed in \cite{Gocksch:1988iz} and recently discussed in \cite{Anagnostopoulos:2001yb,Fodor:2007vv,Langfeld:2014nta,Gattringer:2015lra,Gattringer:2019khb}) with the LLR algorithm \cite{Langfeld:2015fua,Langfeld:2012ah}, as well as proposing a method for the evaluation of the observables in the DoS (density of states) formalism.

At the core of the density of states method is the definition of the DoS function as
\begin{equation}\label{eq:dos_def}
\rho(s) = \int {\cal D}\phi \ \delta(s-S^I[\phi]) \ e^{-S^R[\phi]}
\end{equation}
so that the partition function can be obtained from a 1-dimensional integration
\begin{equation} \label{eq:part_dos}
Z(\mu)= \int \ \rho(s) \ e^{-i\mu s} \ \dd s.
\end{equation}

The severity of the sign problem can then by quantified by the vacuum expectation value of the phase factor
\begin{equation} \label{eq:mv_phase}
\langle e^{i \varphi} \rangle_{pq} =
\frac{Z}{Z_{pq}} = 
\frac{\int \rho(s) \cos(\mu s) \dd s}{\int \rho(s) \dd s} =
e^{- \ V \ \Delta F}.
\end{equation}
A precise evaluation of the integral in the numerator is crucial to obtain reliable results for the free energy $\Delta F$ as the phase factor is exponentially suppressed with V, as shown in Fig.\ref{fig:ph_comp}, making the evaluation of $\rho$ over multiple orders of magnitude fundamental.

In this work we will study the relativistic Bose gas, described on a four dimensional euclidean lattice by the following action
\begin{equation} \label{eq:bose_disc}
S = \sum_x \bigg[ \left(2d+m^2\right) \phi_x^*\phi_x 
 + \lambda\left( \phi_x^*\phi_x\right)^2
- \sum_{\nu=1}^4\left(  \phi_x^* e^{-\mu\delta_{\nu,4}} \phi_{x+\hat\nu} 
+ \phi_{x+\hat\nu}^* e^{\mu\delta_{\nu,4}} \phi_x \right)
\bigg],
\end{equation}
where, by splitting the field into its real and imaginary part, $\phi_x = \phi_{1,x} + i \phi_{2,x}$,
we can separate the real and imaginary part of the action,
\begin{equation}\label{eq:bose_disc_exp}
\begin{aligned}
S 	&= 	S^R+ i \sinh(\mu) S^I \\
S^R	&=	\sum_x\bigg[ \frac{1}{2}\left(
 		2d+m^2\right) \phi_{a,x}^2
 		+ \frac{\lambda}{4}\left(\phi_{a,x}^2\right)^2
 		- \sum_{i=1}^3 \phi_{a, x}\phi_{a, x+\hat i}
 		-\cosh(\mu) \ \phi_{a, x}\phi_{a, x+\hat 4}\bigg]\\
S^I	&=	\sum_x \ \varepsilon_{ab}\phi_{a, x}\phi_{b, x+\hat 4}.
\end{aligned} 
\end{equation}
\section{LLR in a nutshell}
The LLR algorithm is an algorithm that allows us to reconstruct the DoS of continuous systems over several orders of magnitude. Inspired by the successfull Wang-Landau approach to systems with a discrete energy spectrum \cite{Wang:2000fzi}, it is implemented through the following steps:
\begin{enumerate}[noitemsep,topsep=4pt]
\item Divide the complex action domain in N intervals of width $\Delta$
\item For each interval approximate the density of states $\rho$ as
\begin{equation} \label{eq:dos_pw}
\hat{\rho}_k(s) = C_k \exp\left( a_k ( s - S_k^I) \right)
\
\text{with}
\
a_k=\pdv*{\log\rho}{s}\vert_{s=S_k^I}
\end{equation}
\item Obtain $a_k$ as the root of the stochastic equation
\begin{equation}\label{eq:llr}
\langle\langle \Delta S \rangle\rangle_k(a)=
\int_{S_k^I - \Delta/2}^{S_k^I + \Delta/2} \rho(s) \ (s - S_k^I) \ e^{-a ( s - S_k^I)} \ \dd s = 0
\end{equation}
using the Robbins Monro iterative method \cite{Robbins:1951} 
\begin{equation} \label{eq:rm}
a^{(n+1)} =
a^{(n)} + \frac{12 \ \langle\langle \Delta S \rangle\rangle_k (a^{(n)})}{(n+1) \ \Delta^2}
\quad , \quad
\lim_{n\to\infty} a^{(n)} = a_k.
\end{equation}
\end{enumerate}
Using this procedure we are able to sample the $\log \rho$ derivative virtually at any value of the imaginary part of the action. It is important to note that the statement $a^{(\infty)} = a_k$ is true only in the limit of $\Delta \rightarrow 0$. However, in this limit the size of the Robbins Monro correction factor will blow up, making the numerical calculations very inefficient. By evaluating the leading order correction to Eq.\eqref{eq:llr} due to the higher derivatives of the density of states, here defined as $\rho(s) = e^{f(s)}$, it is possible to give an estimate of the intrinsic bias of the LLR method as
\begin{equation}
\text{bias}=a_{\text{biased}}-a_k= \frac{f^{(3)}(S^I_k)}{40} \Delta^2 + \order{\Delta^4}.
\end{equation}
Comparing this quantity with the statistical uncertainty of the LLR simulations it is possible to choose a value of $\Delta$ that minimizes the statistical error while keeping the bias under control. For details see \cite{Francesconi:2019nph}.

\subsection{Density reconstruction}
Having obtained a sampling of the $\log \rho$ derivative, the density of states can be reconstructed straightforwardly following Eq.\eqref{eq:dos_pw}. This leads to a piecewise and continuous definition of $\rho$ that is proved to have a constant relative error over the entire range taken into consideration. However, this reconstruction is not suited for the problem at hands as it introduces discretization error that prevent the precise evaluation the oscillatory integral. To overcome this issue we use the polynomial fit approach, where the LLR results are fitted to a polynomial $p_l(s) = \sum_{i=1}^{l} \ c_{(2i-1)} \ s^{2i-1}$, where only odd powers of $s$ are taken into consideration due to the symmetry properties of the DoS $\rho(s) = \rho(-s)$. The density of states resulting from the polynomial fitting can be expressed as
\begin{equation} \label{eq:dos_fit}
\rho_{\text{fit}(l)}(s) =
\exp{\int_0^s p_l(x) \ \dd x} =
\exp{\sum_{i=1}^l \frac{c_{(2i-1)}}{2i} \ s^{2i}},
\end{equation}
where we are normalising the DoS to have $\rho_{\text{fit}(l)}(0)=1$ as $p_l(0)=0$. For a detailed description of the DoS reconstruction procedure we invite again the reader to refer to \cite{Francesconi:2019nph}.

\section{Free energy}
\begin{figure}[b]
\centering
\includegraphics[width=0.49\textwidth]{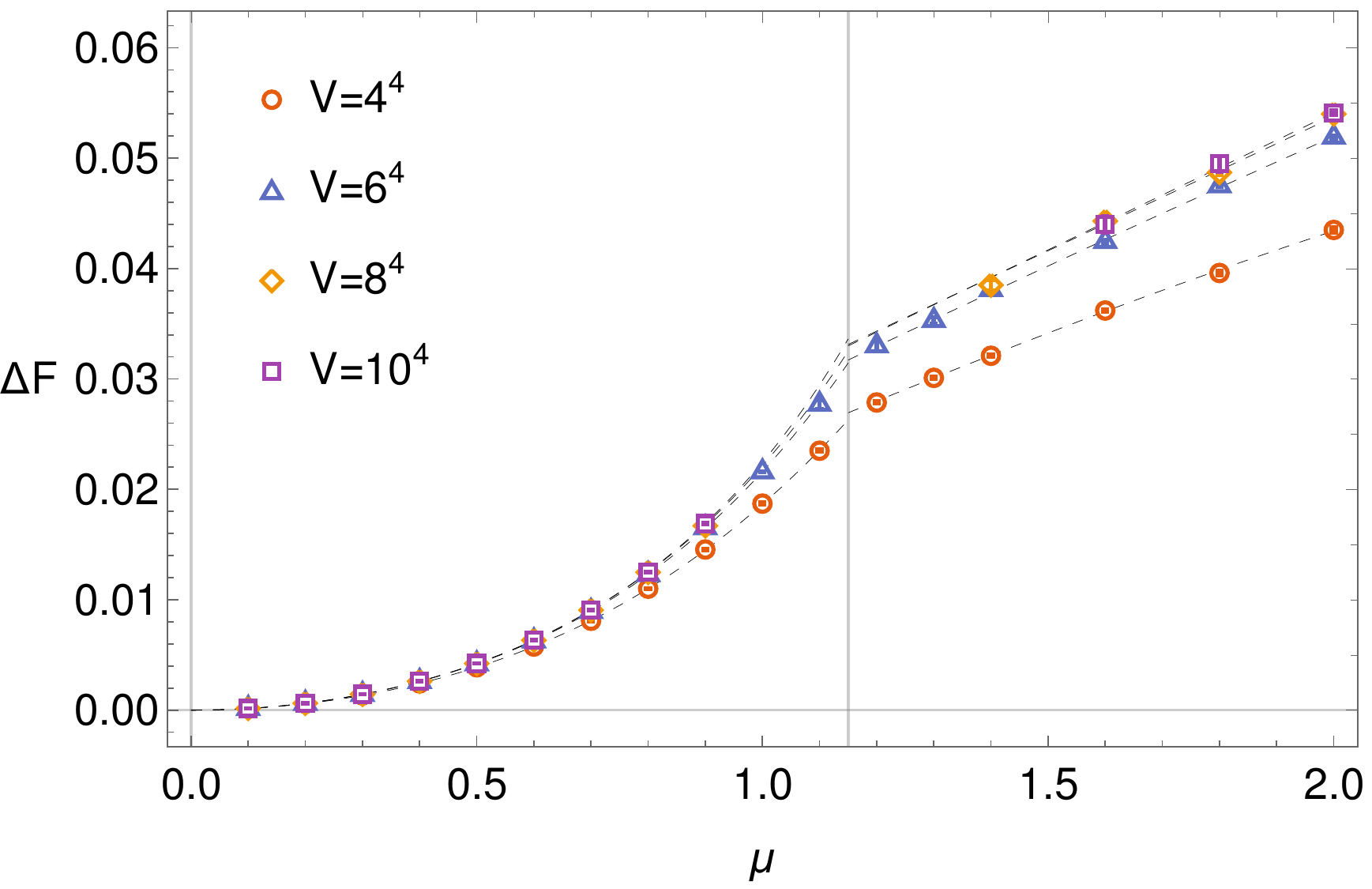}
\includegraphics[width=0.49\textwidth]{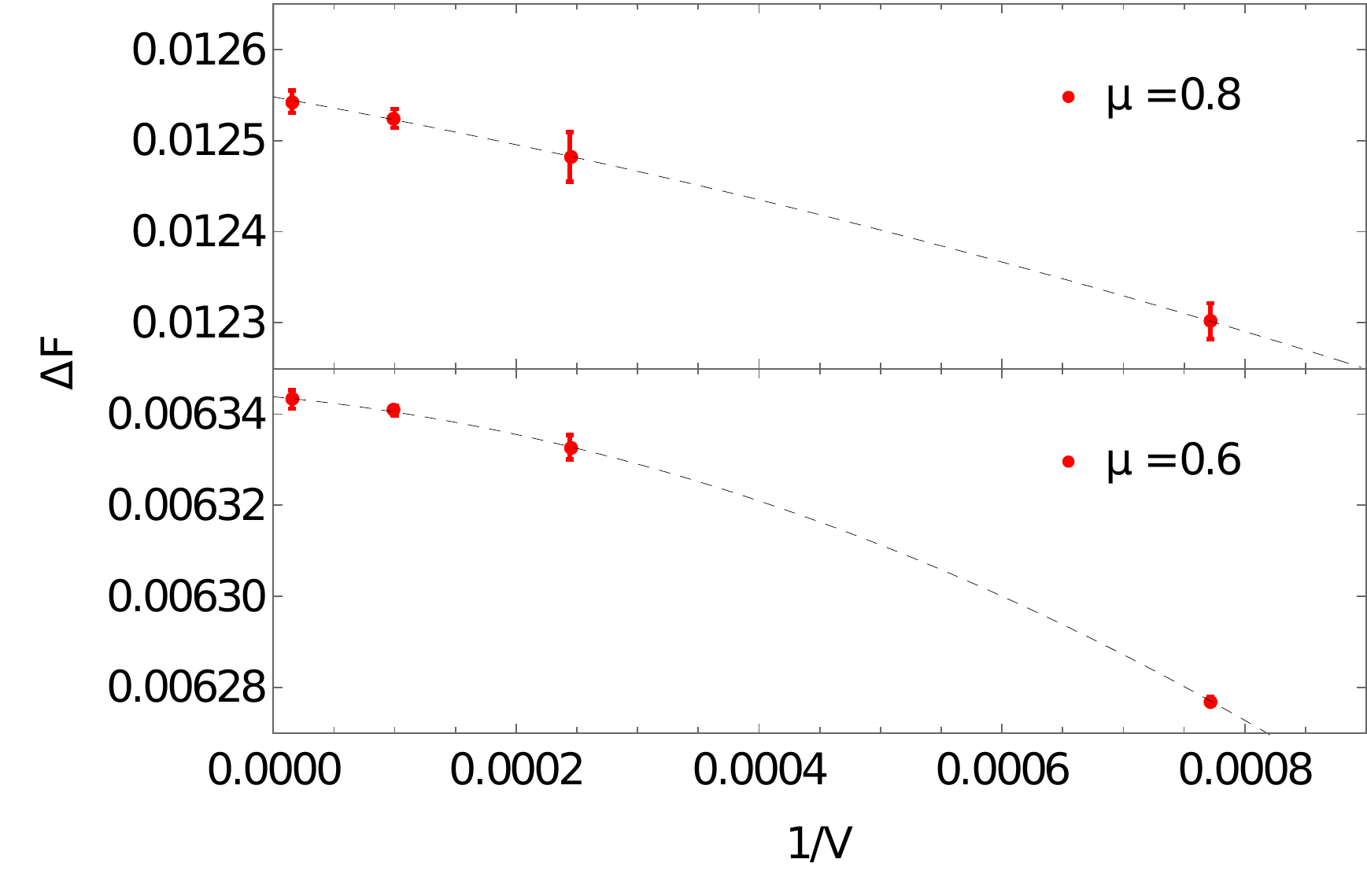}
\caption{\textbf{Left} Free energy difference for chemical potential ranging from 0 to 2.0 and for volumes $4^4$, $6^4$, $8^4$ and $10^4$; the vertical line is the critical value $\mu_c=1.15$ obtained via mean field calculations. \textbf{Right} Infinite volume scaling for the free energy difference in the low density phase up to volumes $V=16^4$, the lines are fits to the data of the form $\Delta F = \Delta F_{V=\infty} + a/V + b/V^2$.}
\label{fig:df}
\end{figure}
With this fitting procedure we can integrate Eq.\eqref{eq:mv_phase}. The results that we are going to show come from a bootstrap analysis where resampled sets of all the $a_k$ values are obtained from an initial sample coming from LLR simulations, the integration is then computed for each polynomial order. The final result is then taken according to the results of the $\chi^2$ analysis. With this procedure we have been able to obtain the results shown in Fig.\ref{fig:df}. In particular, on the left we show the results over a wide range of values of the chemical potential extending in the two known phases of the relativistic Bose gas. The two phases are clearly distinguishable as the behaviour changes qualitatively. It is important to note that our method fails to integrate Eq.\eqref{eq:mv_phase} for volumes bigger than $6^4$ close to $\mu_c$. In this region the behaviour of the $a_k$ requires a too high polynomial order introducing numerical instabilities in the integration. On the right hand side instead we show two examples of infinite volume scaling in the low density region ($\mu < \mu_c$) for volumes up to $16^4$. Notably our results are precise enough to catch also $1 / V^2$ corrections.
\section{General observables}
\begin{figure}[b]
\centering
\includegraphics[width=0.49\textwidth]{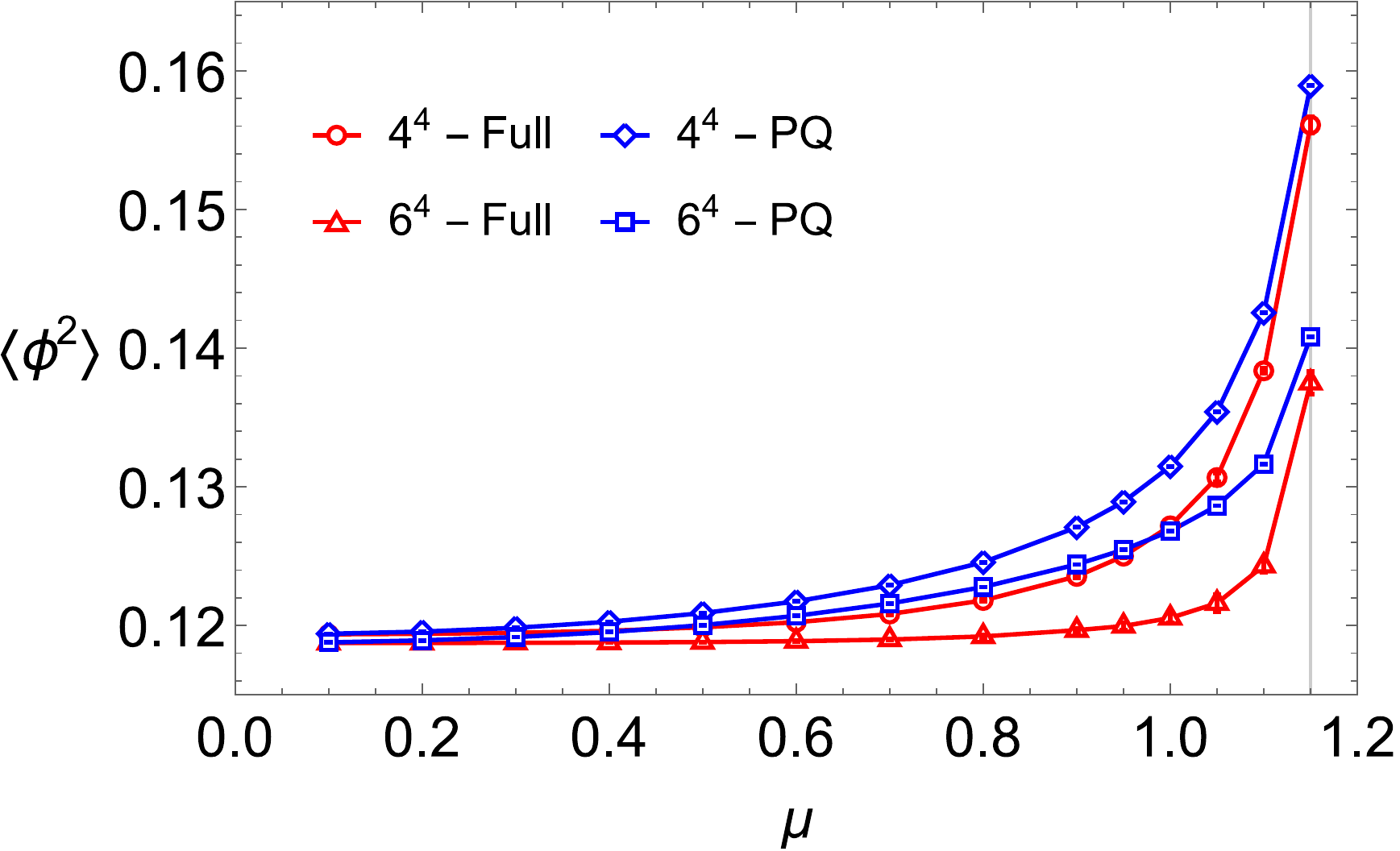}
\includegraphics[width=0.49\textwidth]{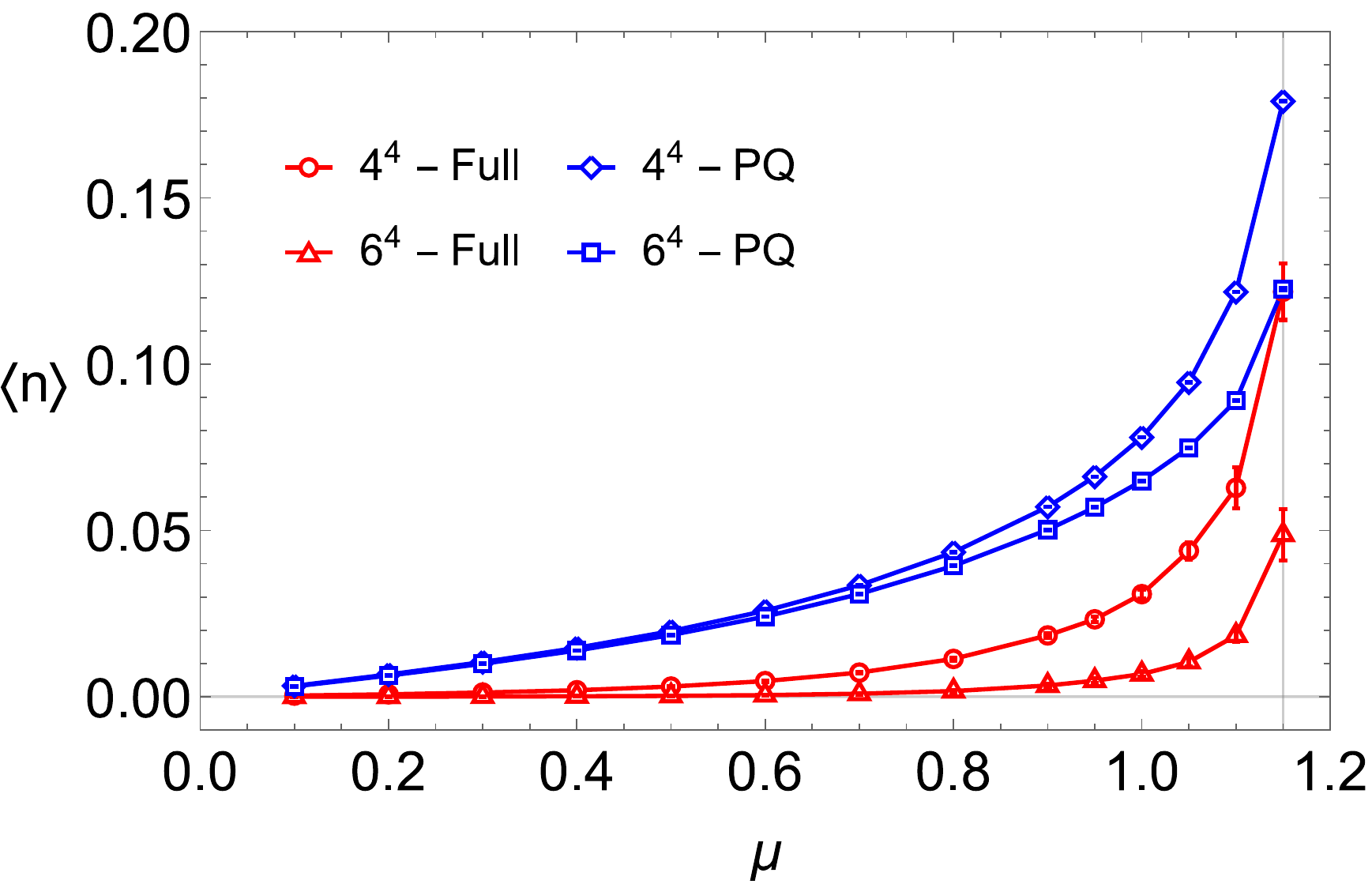}
\caption{Expectation values of the density $\langle n \rangle$ (right) and $\langle \phi^2 \rangle$ (left) for values of the chemical potential ranging from zero to $\mu_c$ on lattices of size $4^4$and $6^4$.}
\label{fig:obs}
\end{figure}
As we have shown, with the density of states approach and the LLR method it is possible to evaluate the phase factor of systems with a complex action over hundreds of orders of magnitude where reweighting techniques fail to obtain reliable results for systems affected by the sign problem. For this reason we turn our attention on the evaluation of general observables (i.e. not limiting ourself to observables that depend only on the imaginary part of the action) in the framework of the density of states approach.

To do so we start with the usual formulation of vacuum expectation values for an observable $X[\phi]$ as path integral over the field variables. We then introduce the same idea of the DoS approach dividing the integration using the delta function on the imaginary action domain and finally introducing the density of states function as defined in Eq.\eqref{eq:dos_def} as shown in the following equations.
\begin{equation}
\begin{aligned}
\langle X \rangle &= \frac{1}{Z} \int \mathcal{D}[\phi] \ X[\phi] \ e^{-S^R[\phi]} \ e^{-i \ \mu \ S^I[\phi]}\\
&= \frac{1}{Z}\int \dd{s} \ e^{-i \ \mu s} \ \rho(s) \frac{\int \mathcal{D}[\phi] \ X[\phi] \ e^{-S^R[\phi]} \ \delta \left( s - S^I[\phi] \right)}{\int \mathcal{D}[\phi] e^{-S^R[\phi]} \ \delta \left( s - S^I[\phi] \right)}\\
&= \frac{1}{Z}\int \dd{s} \ e^{-i \ \mu s} \ \rho(s) \ \tilde{X}(s).
\end{aligned}
\end{equation}
In the last step we have introduced $\tilde{X}(s)$ as the phase-quenched vacuum expectation value of the observable $X[\phi]$ over all the configurations with $S^I[\phi] = s$. The direct evaluation of such observables is impossible. However, with the LLR algorithm we have already defined in Eq.\eqref{eq:llr} a close relative to this expectation value. In particular, we have that $\tilde{X}(s) = \langle\langle X \rangle\rangle_s + \mathcal{O}(\Delta^2)$. With this we can implement efficiently the evaluation of the observables: once the LLR algorithm has obtained a sufficiently accurate estimate of $a_k$ we keep updating the field configuration without updating the reweighting parameter $a$ and proceed to measure the observables of interest as one would do in a standard Monte Carlo simulation. The integration will then proceed as previously described, with the addition of having one extra function describing also the behaviour of the observable in the imaginary action domain.

As shown in Fig.\ref{fig:obs}, where we have plotted the result of a bootstrap analysis, the evaluation of observables using this method is accurate and a clear difference is visible between the full and the phase-quenched expectation values: while the latter clearly depend on $\mu$, in the low density phase the former shows little or no dependence on $\mu$, an expected feature known as Silver Blaze phenomenon. Even though the results shown here are limited to small volumes simulations, it is important to note that the phase factor is already exponentially suppressed getting down to $\mathcal{O}(10^{-20})$ for $V=6^4$ at $\mu=\mu_c$, a region inaccessible via standard Monte Carlo simulations.

\section{DoS of the Thirring model}
\begin{figure}[b]
\centering
\includegraphics[width=0.8\textwidth]{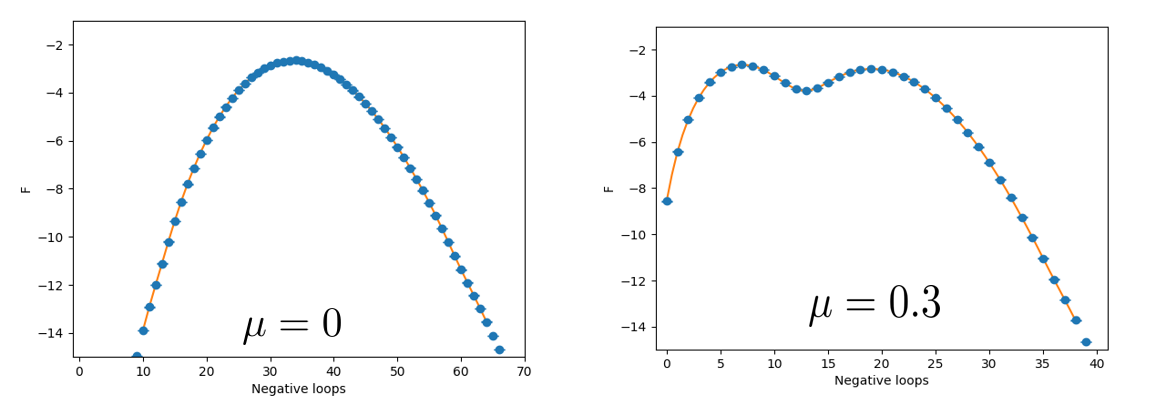}
\caption{Result of a Wang Landau simulation on a $128 \times 128$ lattice with $U=0$ and $m=0.1$.}
\label{fig:thirring}
\end{figure}
Lastly, we present some preliminary results on the evaluation of the density of states for fermionic systems. We study the 2D Thirring model in the world line representation (see \cite{Chandrasekharan:2009wc,Ayyar:2017xmi} for a detailed description) described by the following partition function
\begin{equation}
\begin{aligned}
Z &= \int d\bar\chi d\chi e^{-\sum_{x,y} \left( \bar \chi_x D^{KS}_{x,y} - m\delta_{x,y} \right) \chi_y - U \sum_{x,\nu} \bar\chi_x \chi_x \bar\chi_{x+\nu} \chi_{x+\nu}}\\
& = \sum_{[m,d,l]} m^{N_m} U^{N_d} (-1)^{N_{loops}} \prod_{x,\alpha} \left ( e^{\pm\mu \delta_{\pm\alpha,0}} \frac {s_\alpha \eta_{x,\alpha}}2 \right )^{l_{x,\alpha}},
\end{aligned}
\end{equation}
where the sums runs over configurations of monomers $m$, dimers $d$ and directed loops $l$ (the world lines), $D^{KS}(\mu)$ is the staggered fermion matrix, $\eta_{x,\alpha}$ is the staggered phase, $s_\alpha$ is the sign of direction $\alpha$, m the fermion mass and U the four fermoin coupling. Without getting into the details of the simulations, that are outside the scope of the present work, we focus on the fact that the sign problem (expected for $\mu \ne 0$) in this formulation of the theory comes from the presence of positive and negative valued fermionic loops in the evaluation of the determinant. In particular, it is possible to employ the Wang Landau algorithm \cite{Wang:2000fzi} to measure the DoS of the system as a function of the number of negative loops appearing in the evaluation of the determinant. The results of this simulations are shown in Fig.\ref{fig:thirring}, where we have plotted the free energy of each sector (equivalent to $\log \rho$) against the number of negative loops. These preliminary results show a remarkably well behaved DoS, an encouraging feature that is going to be further analysed in future studies.

\section{Conclusions}
In the present work we have described the density of states approach to systems with complex action. We have reviewed the LLR algorithm and presented some recent advances concerning the bias control of the method. The results from these analysis have been presented for the free energy of the relativistic Bose gas showing a clear change in behaviour between the two phases of the system. A new method to evaluate general observables in the DoS approach has been presented, providing an efficient and straightforward implementation of the method in the LLR algorithm. This has enabled us to obtain numerical results for the density and for $\langle \phi^2  \rangle$ in the full theory, which show the expected Silver Blaze phenomenon, in clear disaccord with the phase quenched case. Lastly, we have presented some results on the Thirring model in the world line formalism that could point to future development of the density of states approach for fermionic systems.

\providecommand{\href}[2]{#2}\begingroup\raggedright\endgroup


\begin{thebibliography}{10}

\bibitem{Gattringer:2016kco}
C.~Gattringer and K.~Langfeld, \emph{{Approaches to the sign problem in lattice
  field theory}}, \href{https://doi.org/10.1142/S0217751X16430077}{\emph{Int.
  J. Mod. Phys.} {\bfseries A31} (2016) 1643007}
  [\href{https://arxiv.org/abs/1603.09517}{{\ttfamily 1603.09517}}].

\bibitem{Francesconi:2019nph}
O.~Francesconi, M.~Holzmann, B.~Lucini and A.~Rago, \emph{{Free energy of the
  self-interacting relativistic lattice Bose gas at finite density}},
  \href{https://arxiv.org/abs/1910.11026}{{\ttfamily 1910.11026}}.

\bibitem{Gocksch:1988iz}
A.~Gocksch, \emph{{Simulating Lattice QCD at finite density}},
  \href{https://doi.org/10.1103/PhysRevLett.61.2054}{\emph{Phys. Rev. Lett.}
  {\bfseries 61} (1988) 2054}.

\bibitem{Anagnostopoulos:2001yb}
K.~N. Anagnostopoulos and J.~Nishimura, \emph{{New approach to the complex
  action problem and its application to a nonperturbative study of superstring
  theory}}, \href{https://doi.org/10.1103/PhysRevD.66.106008}{\emph{Phys. Rev.}
  {\bfseries D66} (2002) 106008}
  [\href{https://arxiv.org/abs/hep-th/0108041}{{\ttfamily hep-th/0108041}}].

\bibitem{Fodor:2007vv}
Z.~Fodor, S.~D. Katz and C.~Schmidt, \emph{{The Density of states method at
  non-zero chemical potential}},
  \href{https://doi.org/10.1088/1126-6708/2007/03/121}{\emph{JHEP} {\bfseries
  03} (2007) 121} [\href{https://arxiv.org/abs/hep-lat/0701022}{{\ttfamily
  hep-lat/0701022}}].

\bibitem{Langfeld:2014nta}
K.~Langfeld and B.~Lucini, \emph{{Density of states approach to dense quantum
  systems}}, \href{https://doi.org/10.1103/PhysRevD.90.094502}{\emph{Phys.
  Rev.} {\bfseries D90} (2014) 094502}
  [\href{https://arxiv.org/abs/1404.7187}{{\ttfamily 1404.7187}}].

\bibitem{Gattringer:2015lra}
C.~Gattringer and P.~Torek, \emph{{Density of states method for the
  $\mathbb{Z}_3$ spin model}},
  \href{https://doi.org/10.1016/j.physletb.2015.06.017}{\emph{Phys. Lett.}
  {\bfseries B747} (2015) 545}
  [\href{https://arxiv.org/abs/1503.04947}{{\ttfamily 1503.04947}}].

\bibitem{Gattringer:2019khb}
C.~Gattringer, M.~Mandl and P.~Törek, \emph{{New DoS approaches to finite
  density lattice QCD}},  \href{https://arxiv.org/abs/1911.05320}{{\ttfamily
  1911.05320}}.

\bibitem{Langfeld:2015fua}
K.~Langfeld, B.~Lucini, R.~Pellegrini and A.~Rago, \emph{{An efficient
  algorithm for numerical computations of continuous densities of states}},
  \href{https://doi.org/10.1140/epjc/s10052-016-4142-5}{\emph{Eur. Phys. J.}
  {\bfseries C76} (2016) 306}
  [\href{https://arxiv.org/abs/1509.08391}{{\ttfamily 1509.08391}}].

\bibitem{Langfeld:2012ah}
K.~Langfeld, B.~Lucini and A.~Rago, \emph{{The density of states in gauge
  theories}}, \href{https://doi.org/10.1103/PhysRevLett.109.111601}{\emph{Phys.
  Rev. Lett.} {\bfseries 109} (2012) 111601}
  [\href{https://arxiv.org/abs/1204.3243}{{\ttfamily 1204.3243}}].

\bibitem{Wang:2000fzi}
F.~Wang and D.~P. Landau, \emph{{Efficient, Multiple-Range Random Walk
  Algorithm to Calculate the Density of States}},
  \href{https://doi.org/10.1103/PhysRevLett.86.2050}{\emph{Phys. Rev. Lett.}
  {\bfseries 86} (2001) 2050}
  [\href{https://arxiv.org/abs/cond-mat/0011174}{{\ttfamily
  cond-mat/0011174}}].

\bibitem{Robbins:1951}
H.~Robbins and S.~Monro, \emph{A stochastic approximation method}, {\emph{The
  Annals of Mathematical Statistics} {\bfseries 22} (1951) 400}.

\bibitem{Chandrasekharan:2009wc}
S.~Chandrasekharan, \emph{{The Fermion bag approach to lattice field
  theories}}, \href{https://doi.org/10.1103/PhysRevD.82.025007}{\emph{Phys.
  Rev.} {\bfseries D82} (2010) 025007}
  [\href{https://arxiv.org/abs/0910.5736}{{\ttfamily 0910.5736}}].

\bibitem{Ayyar:2017xmi}
V.~Ayyar, S.~Chandrasekharan and J.~Rantaharju, \emph{{Benchmark results in the
  2D lattice Thirring model with a chemical potential}},
  \href{https://doi.org/10.1103/PhysRevD.97.054501}{\emph{Phys. Rev.}
  {\bfseries D97} (2018) 054501}
  [\href{https://arxiv.org/abs/1711.07898}{{\ttfamily 1711.07898}}].

\end{thebibliography}
\end{document}